\documentclass[useAMS,usenatbib]{mn2e} 
\usepackage{graphicx}

\title[Autocollimating compensator]
{Autocollimating compensator for controlling aspheric optical surfaces }
\author[V. Yu. Terebizh]{V. Yu. Terebizh$^{1,2}$\thanks{E-mail: 
valery@terebizh.ru}\\
$^{1}$Crimean Astrophysical Observatory, Taras Shevchenko National University of Kyiv, Nauchny,
 Crimea 98409, Ukraine\\
$^{2}$Institute of Astronomy RAN, Moscow 119017, Russian Federation}

\begin{document}

\date{Accepted 2014 February 20. Received 2014 February 18;
 in original form 2014 January 26}

\pagerange{\pageref{firstpage}--\pageref{lastpage}} \pubyear{2014}

\maketitle

\label{firstpage}

\begin{abstract}
A compensator (null-corrector) for testing aspheric optical surfaces is 
proposed, which enables {\it i)}~independent verification of optical elements
and assembling of the compensator itself, and {\it ii)}~ascertaining the 
compensator position in a control layout for a specified aspheric surface.
The compensator consists of three spherical lenses made of the same glass. 
In this paper, the scope of the compensator expanded to a surface speed 
$\sim f/2.3$; a conceptual example for a nominal primary of Hubble Space 
Telescope is given. The autocollimating design allows significant reducing 
difficulties associated with practical use of lens compensators. 
\end{abstract}

\begin{keywords}
telescopes
\end{keywords}

\section{Introduction}

As is known, it is easy to control a spherical surface during its making by
examining the image of a point light source disposed at the center of curvature.
If the sphere is perfect, we should see the diffraction Airy pattern; deviations
from the desired surface shape can be evaluated either qualitatively, by using
the Foucault knife-edge test, or quantitatively, by imposing a reference 
wavefront on the studied wave and analyzing the resulting interferogram. 

On the contrary, testing of aspheric surfaces confronts opticians with a 
much more complicated problem. The point is that the spherical wave, being 
reflected from an aspheric surface, skews its original shape and does not 
form any reasonable image. Let us suppose, for example, that we have to
control a surface of revolution of a conic section of diameter $D$ with the
paraxial curvature radius $R_0$ and squared eccentricity $\varepsilon^2$.
Unlike the sphere, a bundle of normals to this surface do not converge in a
single point; the distance $N(y)$ between points of convergence of the paraxial
normals and those to the $y$-zone -- {\it an aberration of normals} -- is
defined by a simple expression: 
\begin{equation}
  N(y) = \varepsilon^2 s(y),
\end{equation}
where $s(y)$ is sagitta for the $y$-zone (we assume, as usual, that the 
$z$-axis is directed along the system's axis of symmetry). E.g., in the case
of the nominal primary mirror of the Hubble Space Telescope (HST) we have
$D = 2400$~mm, $R_0 = 11040$~mm, $\varepsilon^2 = 1.002299$, so the marginal
sagitta $s(D/2) \simeq 65.22$~mm, and the marginal normals aberration 
$N(D/2) \simeq 65.37$~mm. If we put a point light source in the paraxial 
curvature center, then the diameter of the reflected beam in the source 
vicinity exceeds $28$~mm, and the image-based control of the mirror is 
simply out of the question. 

Under these conditions, aspheric surfaces are usually controlled indirectly.
Namely, a studied aspheric is included as a component into some more extent
optical system in such a way that the wavefront emerging the whole system 
became spherical. The easiest way is to compensate the divergence of 
normals to aspheric surface. The corresponding methods, dating back to 
Maksutov (1924, 1932; see Maksutov 1984, p. 237) and \citet{b1}, have now 
become basic at control of astronomical optics. The theory of compensators
is discussed in detail in monographs of Wilson (1999, Sec.~1.3.4) and
Geary (2002, Ch.~35), as well as in numerous papers mentioned there.

Clearly, tight tolerances are inherent to compensators, but not this 
feature lies in the heart of the problem; quite similar tolerances are 
inherent to some other optical systems. Having high enough optical power, a
compensator must bring into a wavefront huge spherical aberration, moreover,
of strictly specified value. {\em Meanwhile, the conventional compensator 
cannot be verified and properly positioned without the use of extraneous 
optics.} Thus, we must distinguish errors of the tested surface from
those of the compensator and auxiliary devices. For this reason, leading
optical workshops use a few compensators of different type, and only at 
coincidence of the analysis results the surface shape can be considered as
ascertained. 

Let us note in this regard that a known fault at figuring the HST primary 
mirror associated with wrong alignment of two-mirror compensator was 
essentially due to neglecting results obtained with the other, two-lens
compensator of \citet{b7}. According to Wilson (1999, p.~85), equally 
significant errors were made at manufacturing of mirrors for large 
ground-based telescopes, but they have not caused wide public response.

The need for reliable inferences forces us to pay special 
attention to possibilities of {\em independent} checking a compensator 
itself and its position in the control scheme. A few layouts partially 
satisfying to these requirements are described in the literature
\citep{b8,b5,b13}. This paper represents an {\em autocollimating 
compensator}, a device that meets requirements mentioned above. As a 
consequence, one can definitely refer just to the aspheric surface all 
defects of the wavefront remaining visible after performance of 
specified procedures. 

A particular scheme of the autocollimating compensator has been proposed 
several years ago \citep{b10} in connection with the prospective renovation
of the G.A.~Shain 2.6-m telescope in the Crimean Astrophysical Observatory 
(CrAO, Ukraine). Then the compensator was manufactured in  CrAO optical 
workshop under the direction of N.V. Steshenko. A more detailed description
of that layout has been recently given by \citet{b12}. It was noted in the 
latter paper that the compensator to a faster aspheric mirror than the
$f/3.85$ G.A.~Shain primary is of considerable interest. Just this case is 
discussed below on an example of the $f/2.3$ HST nominal primary. Apparently,
the most extensive use of fast compensators can be expected in the domain 
of wide-field survey telescopes \citep{b11}. 

\begin{figure}
  \includegraphics[width=84mm]{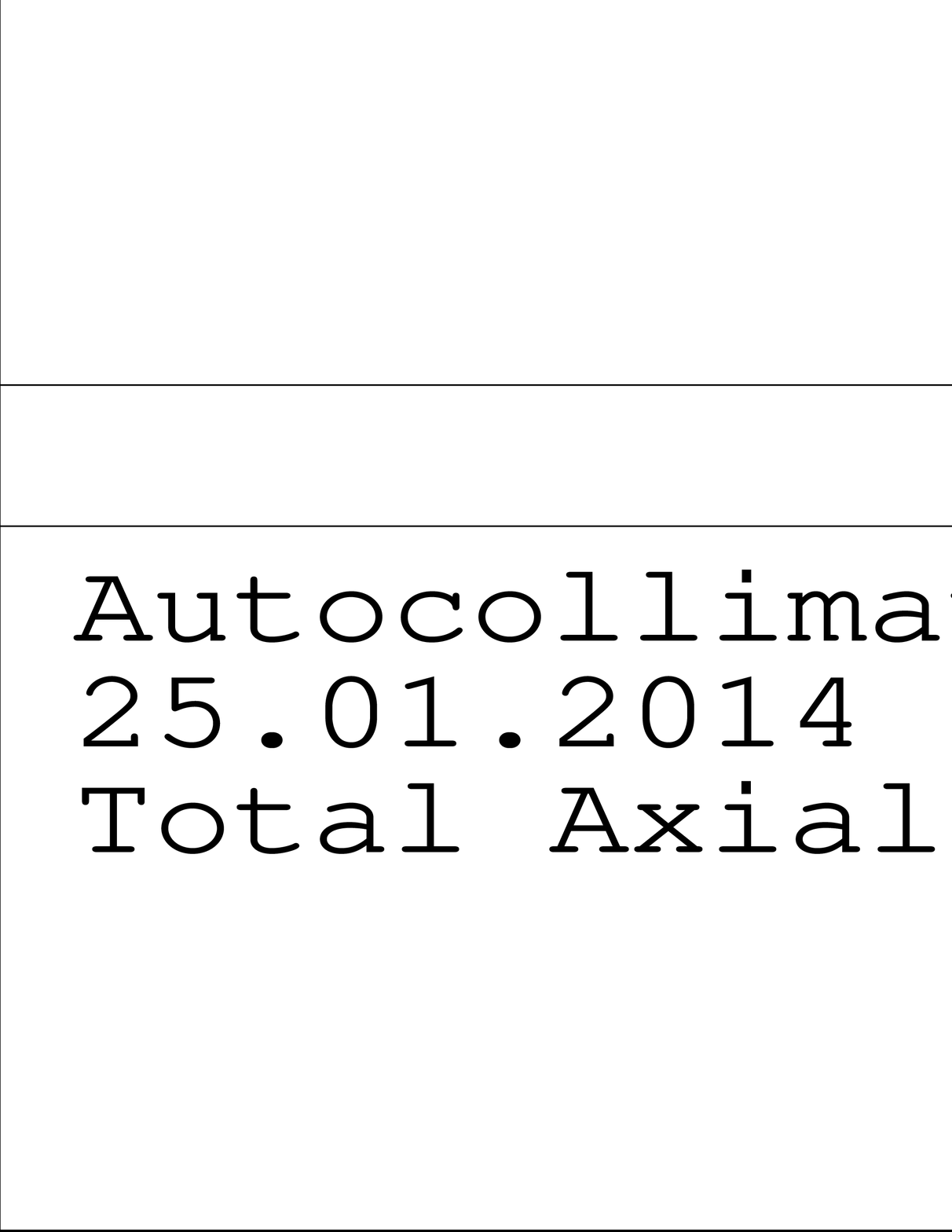}
  \caption{Verification the HST compensator in autocollimating mode.}
  \label{f01}
\end{figure}

\section{Autocollimating mode}

The compensator consists of three spherical lenses made, for the sake of 
simplicity, of the same material (Fig.~1). Choice of glass type is a 
secondary issue; in particular, the compensator for the G.A.~Shain mirror 
has been made of LZOS K8, equal to Schott N-BK7 and Ohara S-BSL7. An 
imaginary compensator for the HST primary discussed here is assumed to be 
made of fused silica, which is more stable to temperature variations. The
light diameter of the largest lens is $111$~mm. The rear surface of the 
third lens is intentionally made flat. 

Fig.~1 and Table~1 give a complete description of the HST compensator in 
a test, autocollimating mode (for brevity, let us call it `A'). If we place
a point light source $S$ at a certain distance $L$ from the vertex of the
first compensator's surface (in our case, $L = 379.062$ mm), then, after 
reflection from the flat surface, light passes the compensator in back 
direction, forming the image in the same place $S$ where the source is 
located. To increase the brightness of reflected light, one can leave the 
flat surface uncoated, or it can be temporarily done specular. By one of 
usual ways the image is put aside; its quality speaks about an 
acceptability of a particular sample of the compensator. 

Obviously, we can reach highest sensitivity at testing by providing the 
diffraction image quality. This requirement is fulfilled for the designed 
HST compensator (Fig.~2); the RMS wavefront error at wavelength
$\lambda = 0.6328\,\mu$m of He-Ne laser is $\lambda/46$. 

\begin{table}
\caption{Compensator in autocollimation mode A.}
\label{symbols}
\begin{tabular}{@{}cccccc}
\hline
 Ele- & $R_0$      & $T$     & Glass  & $D$   & Conic \\
 ment & (mm)       & (mm)    &        & (mm)  &       \\
\hline
Source& $\infty$   & 379.062 & --     & 0     & -- \\
L$_1$ & $-$1200.0  & 16.0    & FS     & 110.0 & 0 \\
      & $-$121.449 & 0       & --     & 110.9 & 0 \\
L$_2$ & 918.520    & 12.0    & FS     & 109.0 & 0 \\
      & $-$245.465 & 6.831   & --     & 108.4 & 0 \\
L$_3$ & $-$139.005 & 12.0    & FS     & 107.6 & 0 \\
      & $\infty$   & 0       & --     & 107.6 & 0 \\
Flat  & $\infty$   & 0       & Mirror & 107.6 & 0 \\
\hline
\end{tabular}

\medskip
 {\small Designations: $R_0$ -- curvature radius, $T$ -- distance to 
 next surface, $D$ -- light diameter, Conic = $-\varepsilon^2$, 
 FS -- fused silica.}
\end{table}

\begin{figure}
  \includegraphics[width=84mm]{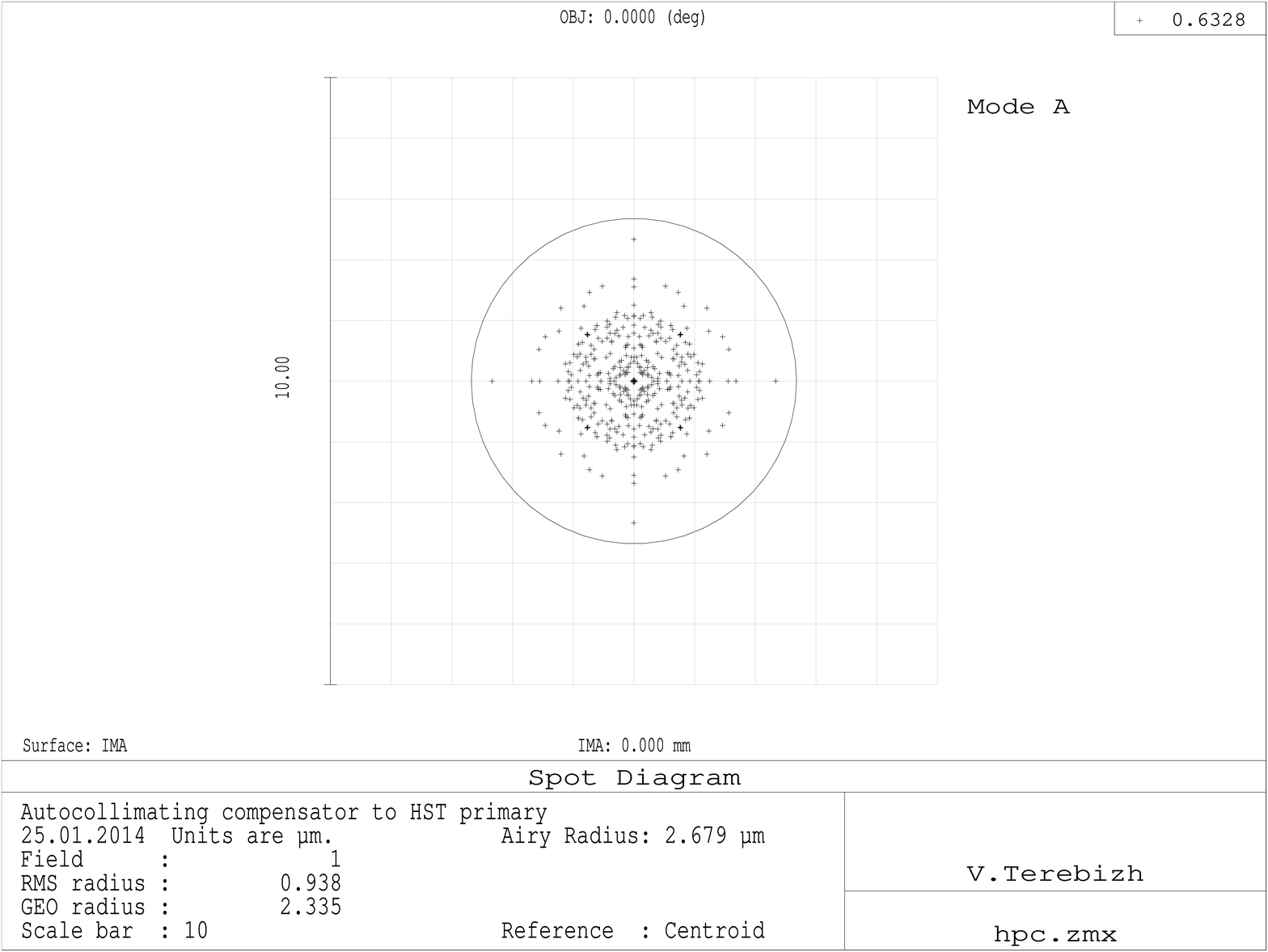}
  \caption{Spot diagram for the HST compensator in autocollimating mode.
  Scale bar corresponds to $10\,\mu$m, the circle of diameter 
  $5.4\,\mu$m -- to Airy disc at $0.6328\,\mu$m.}
  \label{f02}
\end{figure}

Mode A assumes high enough accuracy of the light source arrangement.
As is known \citep{b8}, at use of microscope, longitudinal accuracy
$\delta z$ of position measurement of the source image is determined
by the aperture angle~$u$: 
\begin{equation}
  \delta z \simeq 0.2/u^2,
\end{equation}
where $u$ is measured in radians and $\delta z$ in microns. We have 
for the HST sample $u = 9.4^\circ \simeq 0.16$ radian, so
$\delta z \simeq 8\,\mu$m; this value is within the tolerances.

\section{Control of aspheric surface} 

The basic mode (let us call it `C') corresponds to controlling a 
specified aspheric surface (Fig.~3). Table~2 gives a complete 
description of the particular example for the HST primary mirror. 
Since characteristics of the compensator itself remain identical in 
both the modes A and C, the differences between the Table~1 and 
Table~2 are concerned only with distances from the light source, 
diameters of light beams and inserting of examinee mirror into the 
scheme.

\begin{figure}
  \includegraphics[width=84mm]{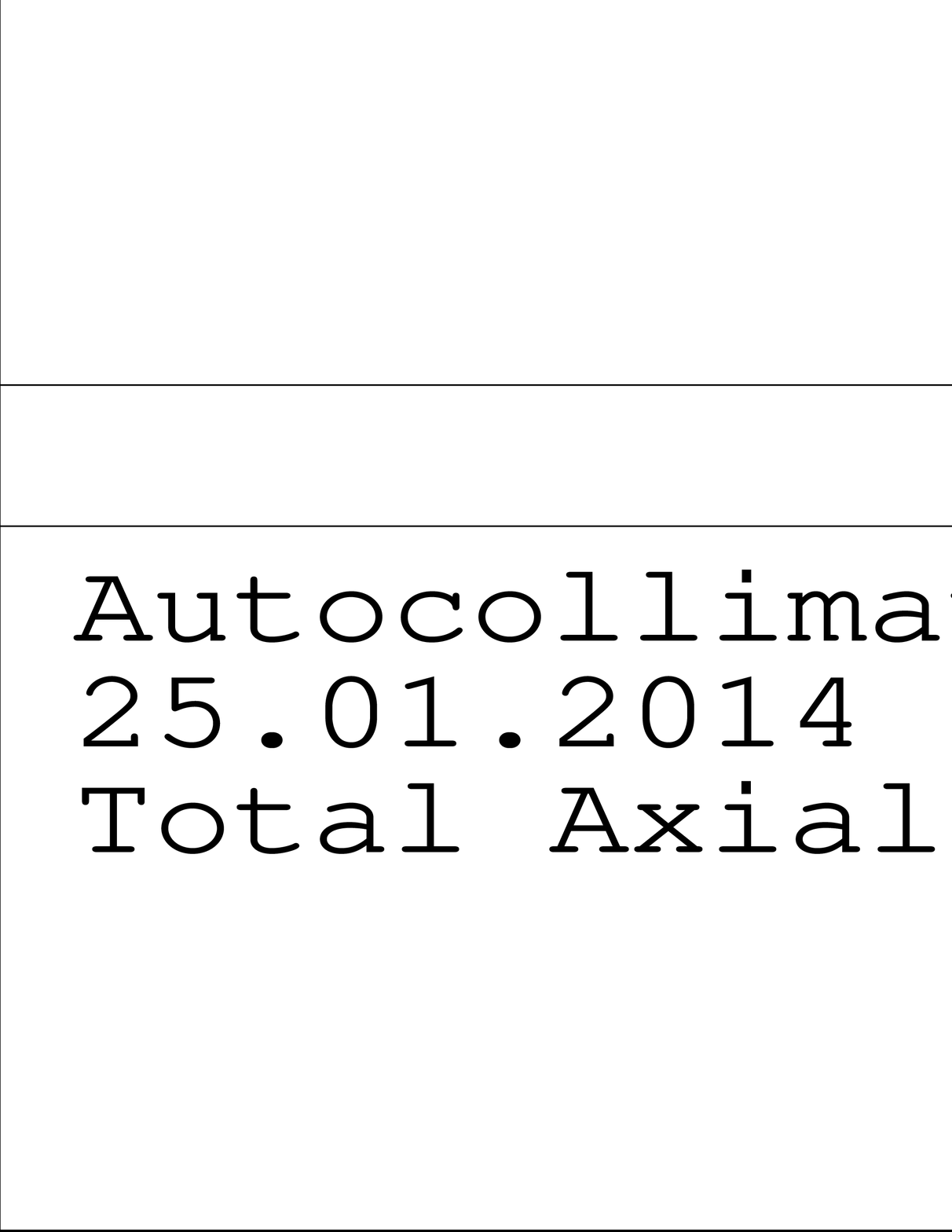} 
  \caption{Compensator in a control mode of the HST primary mirror
  (latter is located far on the right and is not shown in figure).}
  \label{f03}
\end{figure}

One of the essential compensator's features is that the front surface of 
the first lens is concave, and its curvature radius is equal in absolute 
value to a new distance from the light source, in this case $R_1 = -1200.0$~mm
(see Table~2). Thus, considering at first reflection of light from a front 
compensator's surface, we are able to set the spacing between the source and
compensator with optical accuracy, and then, {\em moving the light source 
together with the compensator} and observing the source image in the complete
control scheme, to set also the distance to the vertex of the surface under
control.

\begin{table}
\caption{Compensator for HST primary in control mode C.}
\label{symbols}
\begin{tabular}{@{}cccccc}
\hline
 Ele- & $R_0$      & $T$       & Glass  & $D$   & Conic \\
 ment & (mm)       & (mm)      &        & (mm)  &       \\
\hline
Source& $\infty$   & 1200.0    & --     & 0     & -- \\
L$_1$ & $-$1200.0  & 16.0      & FS     & 110.0 & 0 \\
      & $-$121.449 & 0         & --     & 110.4 & 0 \\
L$_2$ & 918.520    & 12.0      & FS     & 105.2 & 0 \\
      & $-$245.465 & 6.831     & --     & 103.9 & 0 \\
L$_3$ & $-$139.005 & 12.0      & FS     & 102.1 & 0 \\
      & $\infty$   & 11560.11  & --     &  98.9 & 0 \\
  HST & $-$11040.0 &$-$11560.11& Mirror & 2400.0& $-$1.002299 \\
\hline
\end{tabular}

\medskip
{\small The same designations as in Table~1.}
\end{table}

In the C mode, the aperture angle $u = 5.3^\circ = 0.093$ radian, and 
equation (2) estimates the  longitudinal accuracy of setting the light 
source as $\delta z \simeq 23\,\mu$m, which is also within tolerances 
for the complete control scheme.

\begin{figure}
  \includegraphics[width=84mm]{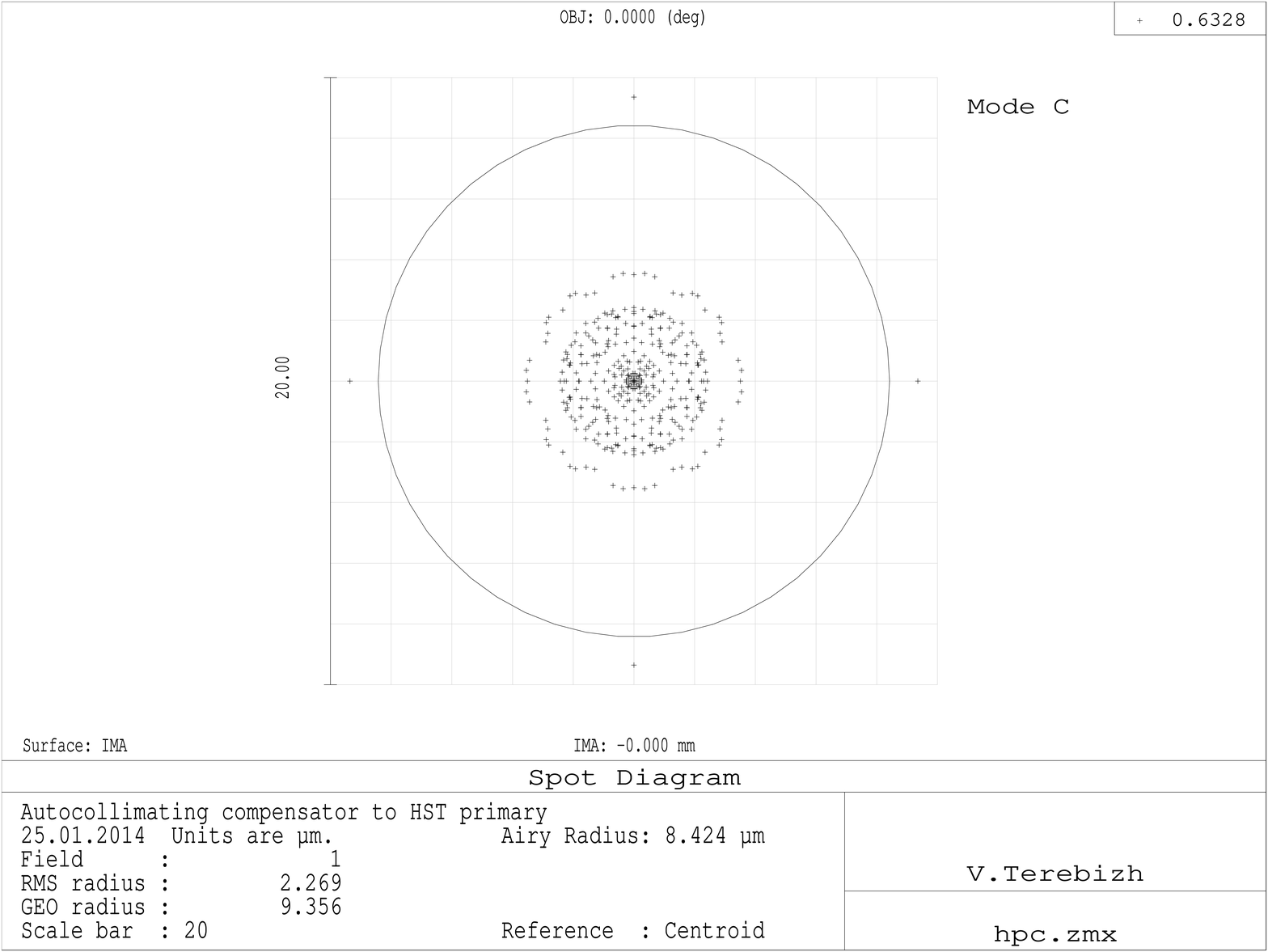} 
  \caption{Spot diagram of the compensator in a control mode with HST primary.
  Scale bar corresponds to $20 \mu$m, the circle of diameter $16.8\,\mu$m --
  to Airy disc at $0.6328\,\mu$m.}
  \label{f04}
\end{figure}

As one can see from Fig.~4, at appropriate placing of the compensator it
provides the diffraction image quality of a point light source. In the C
mode, the RMS error of the wavefront is $\lambda/78$ for 
$\lambda = 0.6328\,\mu$m. Therefore, a possible imperfection of the image
should be attributed only to errors of the tested aspheric surface. The 
specific distribution of errors according to types of aberrations is 
defined by the expansion of the wavefront into Zernike polynomials, 
orthogonal at the annular aperture \citep{b6,b3}.

\section{Tolerances}

Since the HST compensator is considered here only as an illustrative 
example, it is inappropriate to discuss a complete set of tolerances
on its parameters, especially as their list includes 66 items. We will 
confine ourselves by pointing the order of magnitude and comparison 
of tolerances with those for the traditional Offner compensator.

As expected, tolerances of the autocollimating compensator are tight but
common to all compensator types. If we limit the wavefront RMS error by
value $\lambda/20$, then radii tolerances are within $\pm(0.05-0.30)$~mm; 
thickness and surface decenter tolerances lie within $\pm(0.01-0.10)$~mm,
while the elements decenter tolerances are tighter, of the order of 
several microns. Tolerances on the transverse displacement of the 
compensator as a whole are also tight, $\pm7$~$\mu$m.On the other hand, 
tolerances on the index of refraction and Abbe number are not too hard, 
$\pm0.0001$ and $\pm0.3$, respectively.

As a conventional Offner compensator to the HST primary, we have designed 
the system consisting of two singlet lenses of light diameters 93~mm and
31~mm, both made of fused silica. At the previous upper limit for the RMS
wavefront error, all tolerances have the same order of magnitude as the 
specified above, except the noticeably tighter tolerances for radii of 
curvature and refractive index. Perhaps, the reason for more strict
tolerances is that Offner lenses have higher optical power.

Certainly, an advanced compensator can be used as well, e.g., the 
two-mirror Offner null-corrector with an additional field lens that was
actually used for the HST. But, as Wilson (1999, p.~83) noted, ``Such a 
2-mirror Offner compensator may be considered as the ultimate in null
system technology. However, the problems of manufacture and, above all, 
adjustment to correct position remain.''

Let us note that in the course of a compensator design one can
adjust values of all radii to a standard grid of test plates, so the 
implementation of this part of requirements will not be too awkward. 

Generally speaking, the control with a compensator is possible even 
beyond prescribed tolerances, but in that case we need to know the 
actual values of the compensator's parameters and its actual position
in order to account this information at analysis of interferogram
as the inverse problem \citep{b9}.

\section{Concluding remarks} 

As we see, the proposed three-lens compensator has two core features:
\begin{itemize}
  \item There is a position of a point light source, at which light twice 
passes through all lenses and forms a diffraction image at the source 
position. It allows us to find small deviations from nominal characteristics
not only of the lenses, but also their mutual positions. 
  \item The radius of curvature of the first compensator's surface is equal 
in the absolute value to the distance of the compensator from the light 
source in a control mode. This feature allows us to set correctly both the 
light source and compensator.
\end{itemize}

It is worthy to mention, as an additional feature, that all the lenses are
rather small, so one can choose a homogeneous piece of glass of which lenses 
will be made. Besides the specified, there are some other features of 
individual samples. In particular, due to moderate speed of G.A.~Shain 
primary mirror, $f/3.85$, it was possible to reduce the compensator diameter
to $70$~mm and make some of its curvature radii identical.

The autocollimating compensator can be considered as the 
development of a three-lens design by \citet{b8} intended for the control 
of the 6-m BTA primary mirror. In the Puryaev's compensator, the `setting' 
reflection of light occurs not from the first surface, as in our design, 
but from the front surface of the second lens; this feature brings an evident
uncertainty, because the error in distance of the light source can be 
balanced by the deviation of geometrical parameters of first two lenses or 
by the refraction index of the first lens. Eventually, the basic difference
of these two schemes is that our design allows the self-test both in the 
assembled and control states.

\section*{Acknowledgments}

I thank T.A.~Khan, D.A.~Kononov, A.F.~Lagutin, Yu.A.~Petrunin and 
N.V.~Steshenko for useful comments to the paper.

\end{document}